\documentclass[journal]{IEEEtran}

\usepackage{amsmath}   
\usepackage{amssymb}

\RequirePackage{amsthm,amsmath,amssymb,amsfonts,graphicx, subfigure}
\RequirePackage[colorlinks]{hyperref}

\theoremstyle{plain} 
\newtheorem{thm}{Theorem}[section]
\newtheorem{prop}[thm]{Proposition}
\newtheorem{cor}[thm]{Corollary}
\newtheorem{lem}[thm]{Lemma}

\newtheorem{conj}[thm]{Conjecture}

\theoremstyle{remark} 
\newtheorem{rmk}[thm]{Remark}

\newcommand{\bc}{\begin{center}}
\newcommand{\ec}{\end{center}}
\newcommand{\bt}{\begin{tabular}}
\newcommand{\et}{\end{tabular}} 
\newcommand{\bea}{\begin{eqnarray}}
\newcommand{\eea}{\end{eqnarray}}

\newcommand{\ba}{\begin{array}}
\newcommand{\ea}{\end{array}}

\def\be{\begin{eqnarray}}
\def\ee{\end{eqnarray}}
\def\ben{\begin{eqnarray*}}
\def\een{\end{eqnarray*}}

\newcommand{\levy}{\mbox{L\'evy }}

\newcommand{\ra} {\rightarrow}

\newcommand{\nth}{\frac{1}{n}}

\newcommand{\RL}{{\mathbb R}}

\newcommand{\IN}{{\mathbb Z}}

\newcommand{\Nat}{\mathbb{N}}

\newcommand{\COV}{\mbox{\rm Cov}}

\def\sq{$\Box$}

\def\qed{\ifmmode\sq\else{\unskip\nobreak\hfil
\penalty50\hskip1em\null\nobreak\hfil\sq
\parfillskip=0pt\finalhyphendemerits=0\endgraf}\fi\par\medbreak}

\def\tr{{\rm tr\, }}

\newsavebox{\junk}
\savebox{\junk}[1.6mm]{\hbox{$|\!|\!|$}}

\def\det{{\mathop{\rm det}}}
\def\limsup{\mathop{\rm lim\ sup}}
\def\liminf{\mathop{\rm lim\ inf}}

\def\half{{\mathchoice{\textstyle \frac{1}{2}}%
{\frac{1}{2}}%
{\hbox{\tiny $\frac{1}{2}$}}%
{\hbox{\tiny $\frac{1}{2}$}} }}

 \def\eq#1/{(\ref{#1})}

\def\eq#1/{(\ref{e:#1})}

\newcommand{\lam}{\lambda}
\newcommand{\mean}{\text{mean}}
\newcommand{\mode}{\text{mode}}

\def\E{{\bf E}}
\def\R{\RL}

\def\k{{\kappa}}

\def\phi{\varphi}

\newcommand{\eqD}{\mbox{$ \;\stackrel{\mathcal{D}}{=}\; $}}

\def\bee{\begin{eqnarray*}}
\def\ene{\end{eqnarray*}}

\begin{document}
\title{The entropy per coordinate of a random vector is highly constrained under
convexity conditions}
\author{Sergey~Bobkov and~Mokshay~Madiman,~\IEEEmembership{Member,~IEEE}
\thanks{The authors are listed in alphabetical order of last names.}
\thanks{School of Mathematics, University of Minnesota, 127 Vincent Hall, 
206 Church St. S.E., Minneapolis, MN 55455 USA.
Email: \texttt{bobkov@math.umn.edu}}
\thanks{Department of Statistics, Yale University,
24 Hillhouse Avenue, New Haven, CT 06511, USA.
Email: {\tt mokshay.madiman@yale.edu}}
\thanks{SB was supported by NSF grant DMS-0706866,
and MM was supported by a Junior Faculty Fellowship from
Yale University.}}
%
%
%
\markboth{Submitted to IEEE Transactions on Information Theory}{Bobkov and Madiman}
%



\maketitle

\begin{abstract}
The entropy per coordinate in a log-concave random vector of any dimension
with given density at the mode is shown to have a range of just 1. Uniform distributions on convex bodies
are at the lower end of this range, the distribution with i.i.d. exponentially distributed
coordinates is at the upper end, and the normal is exactly in the middle.
Thus in terms of the amount of randomness as measured by entropy per coordinate,
any log-concave random vector of any dimension contains randomness that differs
from that in the normal random variable with the same maximal density value by at most 1/2.
As applications, we obtain an information-theoretic formulation of the famous
hyperplane conjecture in convex geometry, entropy bounds for certain 
infinitely divisible distributions, and
quantitative estimates for the behavior of the density at the mode on convolution.
More generally, one may consider so-called convex or hyperbolic probability measures
on Euclidean spaces; we give new constraints on entropy per coordinate
for this class of measures, which generalize our results under
the log-concavity assumption, expose the extremal role of multivariate Pareto-type
distributions, and give some applications. 
\end{abstract}


%

\section{Introduction}
\label{sec:intro}

\PARstart{A}{} probability density function (or simply ``density'') $f$ defined on the linear space
$\R^n$ is said to be log-concave if 
\be\label{defn:lc}
f(\alpha x +(1-\alpha)y) \geq f(x)^{\alpha} f(y)^{1-\alpha} ,
\ee 
for each $x,y\in \RL^n$ and each $0\leq \alpha\leq 1$. If $f$ is log-concave, we will
also use the adjective ``log-concave'' for a random variable $X$ distributed according
to $f$, and for the probability measure induced by it. (For discussion of the justification
for such terminology, see the beginning of Section~\ref{sec:cvx}.)
Given a random vector $X = (X_1,\dots,X_n)$ in $\R^n$ with density $f(x)$, 
introduce the entropy functional
$$
h(f)= - \int_{\R^n} f(x) \log f(x)\,dx,
$$
provided that the integral exists in the Lebesgue sense; as usual,
we also denote this $h(X)$. Our main contribution
in this paper is the observation that when viewed appropriately, every 
log-concave random vector has approximately the same entropy per coordinate
as a related Gaussian vector. 

Log concavity has been deeply studied in probability, statistics, optimization and geometry, and there are a number of results
that show that log-concave random vectors resemble Gaussian random vectors. For instance,
several functional inequalities that hold for Gaussians also hold for appropriate subclasses
of log-concave distributions (see, e.g., \cite{BE85, Bob99, BBCG08} for discussion
of Poincare and logarithmic Sobolev inequalities for log-concave measures). 
Observe that this is not at all obvious at first glance-- log-concave
probability measures include a large variety of distributions including the uniform distribution
on any compact, convex set, the (one-sided) exponential distribution, and of course any Gaussian. 
In this note, we give a strong (quantitative) information-theoretic basis to
the intuition that log-concave distributions resemble Gaussian distributions. 


To motivate our main results, we first observe that for (one-dimensional) log-concave random variables $X$,
\be\label{mot:1d}
h(X) \approx \log \sigma,
\ee
where $\sigma$ is the standard deviation of $X$. 
(An exact result to this effect is contained in Proposition~\ref{prop:1d}
and proved in Section~\ref{sec:1d}.)
An upper bound for entropy in terms of 
standard deviation clearly follows from the maximum entropy
property of the Gaussian; so it is the lower bound that
is not obvious here. Thus the property \eqref{mot:1d} may be
viewed as asserting comparability between the entropy of 
a one-dimensional log-concave density and that of a Gaussian
density with the same standard deviation.


Our main purpose in this note is to describe a way to
capture the spirit of the statement \eqref{mot:1d} in the 
setting of (multidimensional) random vectors. 
To describe this extension, 
recall that the $L^{\infty}$ norm of a measurable function $f:\RL^n\ra\RL$ is
defined as its essential supremum with respect to Lebesgue measure,
$\|f\|_{\infty}= {\rm ess\, sup}_x f(x)$.
Throughout this paper, we will write $\|f\| =\|f\|_{\infty}$ for brevity.
Any log-concave $f$ is continuous and bounded on the supporting
set $\Omega = \{x: f(x)>0\}$, so we can simply write $\|f\| = \max_{x \in \Omega} f(x)$.

\begin{thm}\label{thm:gaus-comp}
If a random vector $X$ in $\R^n$ has a log-concave density $f$, let 
$Z$ in $\RL^n$ be any normally distributed random vector with 
maximum density being the same as that of $X$. Then 
\ben
\nth h(Z) -\half \leq\, \frac{1}{n}\, h(X)  \leq\, 
\nth h(Z) +\half.
\een
Equality holds in the lower bound if and only if
$X$ is uniformly distributed on a convex set with non-empty interior.
Equality holds in the upper bound if 
$X$ has coordinates that are i.i.d. exponentially distributed.
\end{thm} 

The observation that it is useful to consider Gaussian comparisons
by matching $\|f\|$ rather than the first two moments
may be considered the key observation of this paper.
Theorem~\ref{thm:gaus-comp} follows easily from the following basic proposition
(both are proved in Section~\ref{sec:proof1}). 

\begin{prop}\label{prop:lc}
If a random vector $X$ in $\R^n$ has  density $f$, then
\ben
\frac{1}{n}\, h(X) \,\geq\, \log\, \|f\|^{-1/n} .
\een
If, in addition, $f$ is log-concave, then
\ben
\nth h(X)  \leq\, 1 + \log\, \|f\|^{-1/n}.
\een
\end{prop} 

Observe that the lower bound here is trivial, since
\ben
h(X)
\geq \int_{\R^n} f(x) \log \frac{1}{\|f\|}\,dx = \log \frac{1}{\|f\|} .
\een
On the other hand, let us point out that 
the upper bound in Proposition~\ref{prop:lc} improves upon the naive Gaussian
maximum entropy bound (obtained without a log-concavity assumption). 
Indeed, if the covariance matrix $R$ of $X$
with entries $R_{ij} = {\rm cov}(X_i,X_j)$ is fixed, then $h(X)$ is 
maximized for the normal distribution. This property leads to the 
upper bound
\be\label{not-great}
\frac{1}{n}\, h(X) \leq C + \log \sigma,
\ee
where $\sigma = \det^{1/n}(R)$ and $C = \log \sqrt{2\pi e}$. 
Now, according to one general comparison principle (stated in Section~\ref{sec:ell}),
in the class of all probability densities, the 
quantity $\sigma\,\|f\|^{1/n}$ is minimized for the uniform distribution 
on ellipsoids. 
This property yields
\be\label{strengthen}
\sigma \geq c\,\|f\|^{-1/n} 
\ee
for some universal constant $c>0$. Hence, modulo the constant $C$,
\eqref{not-great} would indeed be improved if we replace $\sigma$ with $\|f\|^{-1/n}$.


While Proposition~\ref{prop:lc} is already remarkable in its own right,
log-concavity is a relatively strong assumption, and it would be advantageous
to loosen it. Inspired by this objective, one wishes to study more general classes of
probability distributions, satisfying weaker convexity conditions
(in comparison with log-concavity). As a natural generalization, we consider 
probability densities of the form
\be\label{defn:kconc-dens}
f(x) = \varphi(x)^{-\beta}, \quad x \in \Omega,
\ee
where $\varphi$ is a positive convex function on an open convex set
$\Omega$ in $\R^n$. 
To see that this is a natural generalization, observe that
any log-concave density is of this form for any $\beta>0$
since the exponential function composed with a convex function is convex,
and that log-concave distributions have finite moments of all orders,
whereas densities of the form \eqref{defn:kconc-dens} can
be heavy-tailed. For example, the Cauchy distribution on the real line has 
density 
$f(x)= [\pi(1+x^2)]^{-1}=\varphi(x)^{-2}$
with $\varphi$ being convex,
although it is certainly not log-concave.

Another example, which is of significant relevance to our development, is 
the $n$-dimensional Pareto distribution.
For fixed parameters $\beta>n$ and $a>0$, 
this has the density
\be\label{gpd}
f_{\beta,a}(x) = \frac{1}{Z_n(\beta,a)}\, (a + x_1 + \dots + x_n)^{-\beta}, \quad x_i>0 ,
\ee
where $Z_n(\beta,a)$ is the normalizing factor, i.e., 
$$
Z_n(\beta,a)= \int_{\RL_{+}^n} \,\, (a + x_1 + \dots + x_n)^{-\beta} \, dx .
$$ 
(As shown in Lemma~\ref{lem:pareto1},
Pareto distributions with $\beta\leq n$ do not exist, since
$Z_n(\beta,a)$ is finite if and only if $\beta>n$.)

\begin{thm}\label{thm:k-conc-comp}
If a random vector $X$ in $\R^n$ has a density
$f$ of the form \eqref{defn:kconc-dens} with $\beta \geq n+1$, and if $\|f\|$ is fixed,
the entropy $h(X)$ is maximal for the $n$-dimensional Pareto distribution.
\end{thm}

Since $h(X)+\log \|f\|$ is an affine invariant, one may assume $\|f\|=1$
without loss of generality. Also, put for definiteness $a=1$ 
and write $Z(\beta)=Z_n(\beta,1)$, and $X_\beta$ for the random vector  
with density $f_{\beta,1}$. Then Theorem~\ref{thm:k-conc-comp}
may be equivalently written as
\be\label{kconc-rewrite}
h(X) + \log\, \|f\|\, \leq\, h(X_\beta) + \log\, \|f_{\beta,1}\| ,
\ee
Moreover, as shown in the Appendix, 
\ben
\frac{1}{Z(\beta)} = (\beta-1) \dots (\beta - n)=(\beta - 1)_n ,
\een 
where $(b-1)_n=\Gamma(b)/\Gamma(b-n)$ is the $n$-th falling factorial of $b-1$,
and \eqref{kconc-rewrite} takes the form
\be\label{kconc-ub}
h(X) + \log\, \|f\|\, \leq\ \beta \sum_{i=1}^n \frac{1}{\beta - i}.
\ee
Hence we recover Proposition~\ref{prop:lc} in the limit as $\beta \rightarrow +\infty$.

It is convenient, for the sake of comparison with Proposition~\ref{prop:lc},
to write some consequences of Theorem~\ref{thm:k-conc-comp} in the following form.

\begin{cor}\label{cor:beta}
For the range $\beta \geq \beta_0 n$ with fixed $\beta_0 > 1$
(and still for $\beta \geq n+1$), we have
$$
\frac{1}{n}\, h(X)  \leq\, C_{\beta_0} + \log\, \|f\|^{-1/n} ,
$$
where the constant $C_{\beta_0}$ depends on $\beta_0$ only. In fact, 
one may take $C_{\beta_0}=\frac{\beta_0}{\beta_0-1}$.
However, in the larger range $\beta \geq \beta_0 + n$ with fixed 
$\beta_0 \geq 1$,
$$
\frac{1}{n}\, h(X)  \leq\, \log\, \|f\|^{-1/n} + O(\log n) ,
$$
where the $O(\log n)$ term may be explicitly bounded. 
\end{cor}

For the range $\beta \leq n$, it is not possible to control $h(X)$ in 
terms of $\|f\|$. In this case $h(X) + \log\, \|f\|$ may be as large, as 
we wish (which can be seen on the example of the Pareto distribution with 
$\beta \rightarrow n$).
One explanation for this observation could be the fact that
the measures with densities \eqref{defn:kconc-dens} for $\beta \leq n$ 
may not be convex (see 
Remark~\ref{rmk:pareto}), 
or viewed another way that there do not exist Pareto distributions for $\beta \leq n$ (see Lemma~\ref{lem:pareto1}).
Thus, we still have a gap $n < \beta < n+1$, when
Theorem~\ref{thm:k-conc-comp} is not applicable, and we cannot say whether one may bound
$h(X)$ in terms of $\|f\|$.


For ease of navigation, let us outline how this note is organized. In Sections~\ref{sec:1d} and \ref{sec:ell},
we expand on the motivation for considering Proposition~\ref{prop:lc} by 
proving the statements \eqref{mot:1d} and \eqref{strengthen}.

In Section~\ref{sec:proof1}, we prove our main results for log-concave probability measures.
In particular, Proposition~\ref{prop:lc} and Theorem~\ref{thm:gaus-comp} emerge
as consequences of a more general result 
that bounds the R\'enyi entropy of any
order $p\geq 1$ using the maximum of the density. As a corollary of this, 
we also show that any two R\'enyi entropies 
become comparable for the class of log-concave densities. 

In Section~\ref{sec:slice}, we use the preceding development to give a new
and easy-to-state entropic formulation of the famous slicing or hyperplane
conjecture. Indeed, the hyperplane conjecture can be formulated
as a multidimensional analogue of the property \eqref{mot:1d},
different from the multidimensional analogue already represented by Theorem~\ref{thm:gaus-comp}.
Specifically, if $D(f)$ is the ``entropic distance'' of $f$ from Gaussianity (defined precisely later),  
then the property  \eqref{mot:1d} may be rewritten in the form
$0\leq D(f) \leq c$ for some constant $c$ and every  one-dimensional log-concave density $f$, 
whereas the hyperplane conjecture is shown to be equivalent to the statement 
that $D(f) \leq cn$ for some universal constant $c$ and every log-concave density $f$
on $\RL^n$. Furthermore existing partial results on the slicing problem are used
to deduce a universal bound on $D(f)$ for all  log-concave densities $f$
on $\RL^n$, although the dominant term in this bound is $\frac{1}{4} n \log n$ rather than linear in $n$.


Section~\ref{sec:cvx} begins the study of a more general class
of probability measures, the so-called ``convex probability measures''.
Sections~\ref{sec:norms} and \ref{sec:proof2} are dedicated to
proving Theorem~\ref{thm:k-conc-comp} (and Corollary~\ref{cor:beta}); the former describes some
necessary tools including a result on norms of convex functions, 
and we complete the proof in the latter.

Section~\ref{sec:appln} develops several applications-- to entropy rates
of certain discrete-time stochastic processes under convexity conditions,
to approximating the entropy of certain infinitely divisible distributions, 
and to giving a quantitative version of an inequality of Junge 
concerning the behavior of $\|f\|$ on convolution.
We end in Section~\ref{sec:disc} with some discussion.



\section{One-dimensional log-concave distributions}
\label{sec:1d}

\begin{prop}\label{prop:1d}
For a one-dimensional log-concave random variable $X$ with standard deviation $\sigma$, 
\ben
\log(C_0 \sigma) \leq h(X) \leq \log(C_1 \sigma)
\een
for some positive constants $C_0,C_1$. The optimal constant $C_1=\sqrt{2\pi e}$
is achieved for the normal, and the optimal constant $C_0>1/\sqrt{2}$.
\end{prop} 

\begin{proof}
The upper bound holds without the log-concavity assumption,
and is obtained  simply by using the Gaussian entropy.

Since $f$ is log-concave, it is supported
on an interval $(a,b)$ (where $a$ may take the value $-\infty$
and  $b$ may take the value $\infty$), and moreover, it is strictly positive
on this support interval (being of the form $e^{-\varphi}$ with $\varphi$ convex).
If  $F$ is the cumulative distribution function of $f$ restricted  to $(a,b)$, its 
inverse $F^{-1}:(a,b) \rightarrow (0,1)$ is well defined since the positivity of $f$
implies that $F$ strictly increases on 
the support interval. Now consider the function
\ben
I(t) = f(F^{-1}(t)) \, ,  \,\,\, 0<t<1.
\een
In \cite[Proposition A1]{Bob96:1}, it was shown that $f$ is log-concave if
and only if $I$ is positive and concave on (0,1).
Hence for all $t\in(0,1)$, $\half I(t)<\half [I(t)+I(1-t)] \leq I(\half)$, 
so that
\ben
I(t) < 2 I(\half) = 2 f(m) .
\een
Taking the supremum over all $t$, one obtains
\be\label{eq:med-max}
\max_x f(x) < 2 f(m).
\ee

For one-dimensional log-concave densities $f(x)$, it was shown in 
\cite[Proposition 4.1]{Bob99} that 
\be\label{bob99}
\frac{1}{12} \leq \sigma^2 f(m)^2 \leq \half ,   
\ee
where $m$ is the median. 
Combining \eqref{eq:med-max} and \eqref{bob99} gives
\ben
\frac{1}{12} \leq \sigma^2 \max_x f(x)^2 < 2 
\een
\be\label{from-bob99}
\text{or}\quad\quad 
\sigma/\sqrt{2} < \|f\|^{-1} \leq \sqrt{12} \sigma .
\ee
Applying Proposition~\ref{prop:lc}, 
\ben
h(X) \geq \log \|f\|^{-1}    >  \log \sigma -\half \log 2 ,
\een
which is the desired lower bound.
\end{proof}

Even in this one-dimensional setting, the best constant $C_0$
and corresponding extremal situations seem to be unknown; these would be interesting 
to identify. Note that the inequalities in \eqref{bob99} are sharp and are attained for
the uniform and double 
exponential distributions.

In Section~\ref{sec:slice}, we discuss the possible generalization of Proposition~\ref{prop:1d}
to general dimension $n$; this is related to the hyperplane conjecture.


\section{An extremal property of ellipsoids}
\label{sec:ell}

Here we recall the comparison property \eqref{strengthen}, mentioned in Section~\ref{sec:intro},
concerning an extremal property of ellipsoids. It goes back to the work
of D. Hensley (\cite{Hen80}, Lemma 2), who noticed that, if a probability density
$f$ on $\R^n$ is maximized at the origin, then the quantity 
$$
f(0)^{2/n} \int_{\R^n} |x|^2\,f(x)\,dx
$$
is minimized for the uniform distribution on the Euclidean balls centered 
at the origin. More precisely, Hensley considered only symmetric quasi-concave
probability densities $f$, and later K. Ball (\cite{Bal88}, Lemma 6) simplified
the argument and extended this observation to all measurable densities $f$
satisfying $f(x) \leq f(0)$ for all $x$.

One may further generalize and strengthen this result, by applying affine 
transformations to the probability measures $\mu$ with densities $f$. 

\begin{prop}\label{prop:ball}
Put 
$$
L_{\mu,\rho} = \int_{\R^n} \rho(\|f\|^{1/n} |x|)\,f(x)\,dx,
$$
where $\rho = \rho(t)$ is a given non-decreasing function in $t \geq 0$. 
In the class of all absolutely continuous probability measures $\mu$ on $\R^n$, 
the functional $L_{\mu,\rho}$ is minimized, when $\mu$ is a uniform distribution
on a Euclidean ball with center at the origin.
\end{prop}

\begin{proof}
Since $L_{\mu,\rho}$ does not depend on $\|f\|$, we may assume $\|f\|=1$.
Denote by $\lambda$ the uniform distribution
on the Euclidean ball $B(0,r_n)$ with center at the origin and volume one
(so that $\omega_n r_n^n = 1$, where $\omega_n$ is the volume of the unit ball).
We need to show that $L_{\mu,\rho} \geq L_{\lambda,\rho}$.
Since both $L_{\mu,\rho}$ and $L_{\lambda,\rho}$ are linear with respect to $\rho$, 
it suffices to consider the case $\rho = 1_{(r,\infty)}$, the indicator function 
of a half-axis. Then the property $L_{\mu,\rho} \geq L_{\lambda,\rho}$ reads as
\[
\mu\{|x| \leq r\}\, \leq\, \lambda\{|x| \leq r\}\, =\, \bigg\{
\begin{array}{cl}
\omega_n r^n, & {\rm for} \ \ 0 \leq r \leq r_n, \\
           1, & {\rm for} \ \ r>r_n. \\
\end{array}
\]
This inequality is automatically fulfilled, when $r>r_n$. In the other case, 
due to the assumption $f(x) \leq 1$ (almost everywhere), we have
$$
\mu\{|x| \leq r\} = \int_{\{|x| \leq r\}} f(x)\,dx \leq 
\int_{\{|x| \leq r\}} \,dx = \lambda\{|x| \leq r\},
$$
which is the statement.
\end{proof}

As a corollary, we obtain the following observation.

\begin{cor}\label{cor:isoLB}
Let $X$ be a random vector in $\RL^n$ with density $f$ and 
non-singular covariance matrix $R$.
If $\sigma^2 = \det^{1/n}(R)$,
\ben
\sigma \geq c\,\|f\|^{-1/n} 
\een
for some universal constant $c>0$. 
\end{cor}
\begin{proof}
Let us return to the basic case $\rho(t)=t^2$. Thus, the functional
$L_{\mu,\rho} = \|f\|^{2/n} \int |x|^2\,f(x)\,dx$ is minimal for 
$\mu = \lambda$, the uniform distribution on the Euclidean ball $B(0,r_n)$ 
with center at the origin and volume one.
Hence, the same is true for the functionals
$$
\frac{1}{n}\ \|f\|^{2/n} \int |T(x-x_0)|^2\,f(x)\,dx
$$
for any point $x_0$ and any linear map $T:\R^n \rightarrow \R^n$ with
$|{\rm det}\, T| = 1$. Taking for $x_0$ the barycenter or mean of $\mu$,
this functional may be written as
$$
\frac{1}{n}\ \|f\|^{2/n} \tr \COV(TX)
$$
where $\tr \COV(TX)$ denotes the trace of the covariance matrix of $T(X)$.
Minimizing over all $T$'s, the above integral turns into
\be\label{ball-inv}
\|f\|^{2/n}\, ({\rm det}\, R)^{1/n},
\ee
where $R$ is the covariance matrix of $X$. 
This follows from the classical representation  (see, e.g., \cite[Proposition II.3.20]{Bha96:book})
for the determinant of a positive-definite matrix $C$:
\ben
({\rm det}\, C)^{\nth}=\min \bigg\{ \frac{\tr(CA)}{n}: A\geq 0, {\rm det}\, A=1 \bigg\} .
\een

The point is that the quantity 
\eqref{ball-inv} is invariant both under all shifts and all linear transforms of $\R^n$.
In particular, it is constant for the uniform distribution on all ellipsoids, 
which thus minimize \eqref{ball-inv}. Analytically, for any probability density $f$,
$$
\|f\|^{2/n}\, ({\rm det} R)^{1/n} \geq \frac{1}{n}\ \int_{B(0,r_n)} |x|^2\,dx
= \frac{r_n^2}{n+2}= \frac{\omega_n^{-2/n}}{n+2}.
$$
Since $r_n$ is of order $\sqrt{n}$ for the growing dimension $n$, 
the right side is separated from zero by a universal constant.
\end{proof}

In fact, this proof allows us to compute the optimal dimension-free 
constant. Recall that the volume of the unit ball is
$\omega_n= \pi^{n/2}/\Gamma(\frac{n}{2}+1)$.
Restricting  ourselves for simplicity to even dimension
$n$, the optimal dimension-dependent lower bound becomes
\ben
\frac{\omega_n^{-2/n}}{n+2}
= \frac{\pi^{-1}}{(\frac{n}{2})!^{-2/n}} \cdot\frac{1}{n+2} ,
\een 
which by Stirling's approximation is multiplicatively
well-approximated for large $n$ by
\ben\begin{split}
\frac{1}{\pi(n+2)} \bigg[ \sqrt{2\pi\bigg(\frac{n}{2}\bigg)} \cdot \bigg(\frac{n}{2e}\bigg)^{\frac{n}{2}} \bigg]^{\frac{2}{n}} 
= \frac{1}{2\pi e} \cdot \frac{n}{n+2} \cdot (\pi n)^{\nth}  .
\end{split}\een
As $n\ra\infty$ through the subsequence of even numbers, this quantity
converges to $c=(2\pi e)^{-1}$, which is therefore the optimal dimension-free constant.
Observe that when Corollary~\ref{cor:isoLB} is written with this
dimension-free constant, equality is not attained
for any finite dimension $n$ but only asymptotically. 
In Section~\ref{sec:slice}, we give a very simple proof of
Corollary~\ref{cor:isoLB} using entropy that also naturally yields 
the exact dimension-free constant.


\section{R\'enyi entropies of log-concave distributions}
\label{sec:proof1}

Recall the definition of the R\'enyi entropy of order $p$: for $p>1$,
and a random vector $X$ in $\RL^n$ with density $f$,
\ben
h_p(X)= \frac{p}{p-1}\log \frac{1}{\|f\|_p} ,
\een
where 
\ben
\|f\|_p= \bigg(\int_{\RL^n} f^p\,dx\bigg)^{\!1/p}
\een
is the usual $L^p$-norm with respect to Lebesgue measure on $\R^n$.
By continuity, $h_p(X)$ reduces to the Shannon differential entropy
$h(X)$ as $p\ra 1$, and to $h_{\infty}(X)=\log \|f\|^{-1}$ as $p\ra\infty$. 
The definition of $h_p(X)$ continues to make sense
for $p\in (0,1)$ even though $\|f\|_p$ is then not a norm.

\begin{thm}\label{thm:lc}
Fix $p\in (1,\infty)$. If a random vector $X$ in $\R^n$ has  density $f$, then
\ben
\frac{1}{n}\, h_p(X) \,\geq\, \log\, \|f\|^{-1/n} ,
\een
with equality if and only if $X$ has the uniform distribution on any set of positive finite Lebesgue
measure. If, in addition, $f$ is log-concave, then
\ben
\nth h_p(X)  \leq\, \frac{1}{p-1}\log p + \log\, \|f\|^{-1/n} ,
\een
with equality for the $n$-dimensional exponential distribution, concentrated on the positive
orthant with density $f(x) = e^{-(x_1 + \dots + x_n)}$, $x_i > 0$. 
\end{thm} 

\begin{proof}
The lower bound is trivial and holds without any assumption on the
density. 

Let us derive the upper bound for $p>1$. By definition of 
log-concavity, for any $x,y \in \R^n$,
\be\label{lc-defn}
f(tx + sy) \geq f(x)^t\, f(y)^s, \, t,s> 0, t+s = 1.
\ee
Taking the $t$-th root yields
$$
f(tx + sy)^{1/t} \geq f(x)\, f(y)^{s/t}.
$$
Integrating with respect to $x$ and using the assumption that $\int f = 1$, 
we get
$$
t^{-n} \int f(x)^{1/t}\,dx \geq f(y)^{s/t}.
$$
It remains to optimize over $y$'s, so that
$$
\int f(x)^{1/t}\,dx \geq t^{n}\, \|f\|^{s/t}.
$$
Taking $p=1/t$ implies $\int f^p \geq p^{-n} \|f\|^{p-1}$ or
\ben
\|f\|_p^{-1} \leq p^{n/p} \|f\|^{\frac{1-p}{p}} ,
\een
so that
\ben\begin{split} 
h_p(X)  
&\leq \frac{p}{p-1}\log [ p^{n/p} \|f\|^{\frac{1-p}{p}} ] \\
&= \frac{n}{p-1}\log p + \log \|f\|^{-1} .
\end{split}\een
It is easy to check that a product of exponentials is an instance of equality. 
\end{proof}

Thus a maximizer of the R\'enyi entropy of order $p$ under a log-concavity shape constraint
and a supremum norm constraint is the exponential distribution, irrespective of $p$. 
This is not the only maximizer-- indeed, affine transforms with determinant 1 of an
exponentially distributed random vector will also work. 
Let us remark that if one instead imposes a variance constraint, the maximizers of
R\'enyi entropy are Student's distributions as shown by
Costa, Hero and Vignat \cite{CHV03}, which specialize to the Gaussian for $p=1$. 
(See also Johnson and Vignat \cite{JV07} and Lutwak, Yang and Zhang 
\cite{LYZ04, LYZ07a} for additional related results.)


We may now prove some of the results stated in Section~\ref{sec:intro}. 

\begin{proof}[Proof of Proposition~\ref{prop:lc}]
Note that Proposition~\ref{prop:lc} is just a limiting version of Theorem~\ref{thm:lc},
obtained by letting $p\downarrow 1$. However, it is not 
automatic,  
since there exist densities such that $h_p(X)=\infty$
for every $p>1$ but $h(X)<\infty$. (An example of such a density is 
\ben
f(x)=\frac{c}{x\log^3(1/x)} , \,\, 0<x<\half ,
\een
where $c$ is a normalizing constant.)  Note that by L'H\^{o}pital's rule,
what one needs to show is that 
\ben
\lim_{p\downarrow 1}   \,\, -\bigg(\int f^p\bigg)^{-1} \frac{d}{dp} \int f^p
\een
exists and equals $h(X)$. This calls for three limit interchanges,
each of which can be justified by the Lebesgue dominated convergence theorem
if $h_p(X)$ is finite for $p\in (1,2]$.  In our context
of log-concave densities, this is always the case because of Theorem~\ref{thm:lc}
and the boundedness of log-concave densities.
Alternatively, a direct proof of Proposition~\ref{prop:lc} can be 
given similar to that of Theorem~\ref{thm:lc} by integrating \eqref{lc-defn}
with respect to $x$, maximizing over $y$,
and then comparing derivatives in $t$ at $t=1$.
\end{proof}

\begin{proof}[Proof of Theorem~\ref{thm:gaus-comp}]
To see the relationship with the Gaussian, simply observe that the maximum
density of the $N(0,\sigma^2 I)$ distribution is $(2\pi \sigma^2)^{-n/2}$.
(Here as usual, we use $N(\mu,\Sigma)$ to denote the Gaussian distribution with
mean $\mu$ and covariance matrix $\Sigma$.)
Thus matching the maximum density of $f$ and the isotropic normal $Z$ leads
to $(2\pi \sigma^2)^{1/2}=\|f\|^{-1/n}$, and
\ben
\nth h(Z)=\half\log (2\pi e \sigma^2) = \half + \log \|f\|^{-1/n}.
\een
This completes the proof of Theorem~\ref{thm:gaus-comp}.
\end{proof}


Theorem~\ref{thm:lc} also implies that for log-concave random vectors,
R\'enyi entropies of orders $p$ and $q$ are related for any $p,q\geq 1$.

\begin{cor}\label{cor:renyi-compare}
If $X$ has a log-concave distribution on $\R^n$, and $p,q\in [1,\infty]$, then
\ben
\frac{h_p(X)}{n} \leq \frac{\log p}{p-1} + \frac{h_q(X)}{n} .
\een
\end{cor}

Since Theorem~\ref{thm:lc} is just the special case $q=\infty$ of
Corollary~\ref{cor:renyi-compare}, the two statements are mathematically equivalent.

While the preceding discussion relies heavily on the value of the density
at the mode, one can also extract information based on the value at the mean.
Let $g: \RL^n \ra [0,\infty)$ be a log-concave function such that
$\int g \in (0,\infty)$. Let $x_{\mean}$ be the barycenter or mean of $g$. Then
it was shown by Fradelizi \cite{Fra97} that
\ben
\sup_{x\in\RL^n} g(x) \leq e^{n}g(x_{\mean}) .
\een
Combining Proposition~\ref{prop:lc}
with Fradelizi's lemma immediately yields the following corollary.

\begin{cor}
If a random vector $X$ has log-concave density $f$, with mean $x_{\mean}$
and mode $x_{\mode}$, then
\ben
h(X) \in [\log f(x_{\mean})^{-1}-n , \log f(x_{\mode})^{-1}+n ]. 
\een
\end{cor}


\section{An Entropic Formulation of the Slicing Problem}
\label{sec:slice}

The main observation of this section is a relationship between
the entropy distance to Gaussianity $D(f)$ and the isotropic
constant $L_f$ for densities of convex measures. 

For a random vector $X$ with density $f$ on $\RL^n$, 
the relative entropy from Gaussianity $D(f)$ or $D(X)$ is defined by
\ben
\int f(x) \log\frac{f(x)}{g(x)} dx ,
\een
where $g$ is the density of the Gaussian distribution with the same mean
and the same covariance matrix as $X$. If $Z$ has density $g$,
then one may write $D(X)=h(Z)- h(X)$ (see, e.g., Cover and Thomas \cite{CT91:book}).


For any probability density function $f$ on $\RL^n$ with covariance matrix $R$,
define its isotropic constant $L_f$ by
\ben
L_f^2=\|f\|^{2/n} \det^\nth(R) .
\een
The isotropic constant has a nice interpretation for uniform distributions
on convex sets $K$. If one rescales $K$ (by a linear
transformation) so that the volume of the convex set is 1 and the covariance matrix is
a multiple of the identity, then $L_K^2:=L_f^2$ is the value of the
multiple.

Observe that both $D(f)$ and $L_f$ are affine invariants. The following result
relating them may be viewed as an alternative form of Theorem~\ref{thm:gaus-comp}
relevant to matching first and second moments rather than the supremum norm.

\newcommand{\vol}{\text{Vol}}
\begin{thm}\label{thm:D-iso}
For any density $f$ on $\RL^n$,
\ben
\nth D(f) \leq \log [\sqrt{2\pi e} L_f ] ,
\een
with equality if and only if $f$ is the uniform density on some
set of positive, finite Lebesgue measure.
If $f$ is a log-concave density on $\RL^n$, then
\ben
\log \bigg[\sqrt{\frac{2\pi}{e}} L_f \bigg] \leq \nth D(f) ,
\een
with equality if $f$ is a product of one-dimensional 
exponential densities.
\end{thm}

\begin{proof}
Let $X\sim f$ have covariance matrix $R$. If $Z\sim N(0,R)$,
\ben
h(Z)=\half\log [(2\pi e)^n \det(R)] = \frac{n}{2} \log (C\sigma^{2}) ,
\een
where $\sigma^{2}=\det(R)^\nth$ and $C=2\pi e$. Thus
\ben\begin{split}
\nth D(X)&= \frac{h(Z)-h(X)}{n} \\
&\leq \half \log (C\sigma^{2}) - \log \|f\|^{-\nth} \\
&= \half \log [C\sigma^{2}\|f\|^{2/n}] 
= \half \log [C L_f^2] ,
\end{split}\een
and 
\ben\begin{split}
\nth D(X)&= \frac{h(Z)-h(X)}{n} \\
&\geq \half \log (C\sigma^{2}) - \log \|f\|^{-\nth} - 1\\ 
&= \half \log \bigg[\frac{C}{e^2}\sigma^{2}\|f\|^{2/n}\bigg] 
= \half \log \bigg[\frac{2\pi}{e} L_f^2\bigg] ,
\end{split}\een
where the inequalities come from Proposition~\ref{prop:lc}.
\end{proof}

Note that this immediately gives an extremely simple alternate proof 
of Corollary~\ref{cor:isoLB}. Indeed, since $D(f)\geq 0$,
we trivially have 
\ben
\sqrt{2\pi e} L_f \geq 1 ,
\een
which is Corollary~\ref{cor:isoLB} with the optimal dimension-free constant.


On the other hand, whether or not 
the isotropic constant
is bounded from above by a universal 
constant for the class of uniform distributions on convex 
bodies is an open problem that has attracted a lot of attention in the last 20 years.
It was originally raised by J. Bourgain \cite{Bou86} in (a slight variation of) the following form.

\begin{conj}\label{conj:hyp1}[{\sc Slicing Problem or Hyperplane Conjecture}] 
 There exists a universal, positive constant $c$ (not depending on $n$) 
such that for any convex set $K$ of unit volume in $\mathbb{R}^n$,  there exists a hyperplane $H$
such that the $(n-1)$-dimensional volume of the section
$K\cap H$ is bounded below by $c$. 
\end{conj}

There are several equivalent formulations of the conjecture, all of a geometric or functional
analytic flavor. Whereas Bourgain \cite{Bou86} and Milman and Pajor \cite{MP89} looked at aspects of the
conjecture in the setting of centrally symmetric, convex bodies, a popular formulation developed  by Ball \cite{Bal88}
is that the isotropic constant of a log-concave measure in any
Euclidean space is bounded above by a universal constant independent of dimension.
Connections of this question with slices of $\k$-concave measures
are described in \cite{Bob10}.

We will now demonstrate that the hyperplane conjecture has a formulation 
in purely information-theoretic terms. 
It is useful to start by mentioning the following equivalences.

\begin{cor}\label{cor:D-iso}
Let $c(n)$ be any non-decreasing
sequence, and $c'(n)=c(n)+\half\log(2\pi e)$.
Then the following statements are equivalent:
\begin{enumerate}
\item[(i)] For any log-concave density  $f$ on $\RL^n$,
 $L_f \leq e^{c(n)}$.
\item[(ii)] For any log-concave density  $f$ on $\RL^n$,
 $D(f) \leq nc'(n)$.
\item[(iii)] 
 $\sup_f \min_g D(f\|g) \leq nc'(n)$,
 where the minimum is taken over all Gaussian densities on $\RL^n$,
 and the maximum is taken over all log-concave densities on $\RL^n$.
\end{enumerate}
\end{cor}

\begin{proof}
The equivalence of (i) and (ii) follows from
Theorem~\ref{thm:D-iso}, and that of
(ii) and (iii) follows from the easily verified fact that $D(f)=\min_g D(f\|g)$,
where $g$ is allowed to run over all Gaussian distributions.
\end{proof}

Furthermore, the seminal paper of Hensley \cite{Hen80} (cf. Milman and Pajor \cite{MP89})
showed that for an isotropic convex body $K$, and any hyperplane $H$ passing through
its barycenter,
\ben
c_1 \leq L_K \vol_{n-1}(K\cap H) \leq c_2 ,
\een
where $c_2>c_1>0$ are universal constants. Hence the statements
of Corollary~\ref{cor:D-iso}, when restricted to uniform distributions
on convex sets, are also equivalent to the statement that
\ben
\vol_{n-1}(K\cap H)\geq e^{-c(n)} . 
\een
Thus the slicing problem or the hyperplane conjecture is simply 
the conjecture that $c(n)$ can be taken to be 
constant (independent of $n$), in {\it any} of the statements of Corollary~\ref{cor:D-iso}.

\begin{conj}[{\sc Entropic Form of Hyperplane Conjecture}]\label{conj:hyp2}
For any log-concave density $f$ on $\RL^n$ and some universal
constant $c$,
\ben
\frac{D(f)}{n}\leq c .
\een
\end{conj} 

This gives a pleasing formulation of the slicing problem as a statement
about the (dimension-free) closeness of an arbitrary log-concave measure 
to a Gaussian measure. 

Let us give another entropic formulation as a statement about the 
(dimension-free) closeness of an arbitrary log-concave measure 
to a {\it product} measure. 
If $f$ is an arbitrary density on $\RL^n$ and  $f_i$ denotes the $i$-th marginal of $f$,
set 
\ben
I(f)=D(f\|f_1 \otimes f_2 \otimes\ldots\otimes f_n) ;
\een
this is the ``distance from independence'', or the relative entropy of $f$ from
the distribution of the random vector that has the same one-dimensional
marginals as $f$ but has independent components. (For $n=2$, this reduces to the
mutual information.)

\begin{conj}[{\sc Second Entropic Form of Hyperplane Conjecture}]\label{conj:hyp3}
For any log-concave density $f$ on $\RL^n$ and some universal
constant $c$,
\ben
\frac{I(f)}{n}\leq c .
\een
\end{conj} 

\begin{proof}[Proof of equivalence of Conjectures~\ref{conj:hyp2} and \ref{conj:hyp3}]
The following identity is often used in information theory:
if $f$ is an arbitrary density on $\RL^n$ and $f^{(0)}$ is the density 
of some product distribution (i.e., of a random vector with independent components),
then
\be\label{eq:chain}
D(f\|f_0)= \sum_{i=1}^n D(f_i\| f^{(0)}_i) + I(f) ,
\ee
where $f_i$ and $f^{(0)}_i$ denote the $i$-th marginals of $f$ and $f^{(0)}$ respectively.

Now Conjecture~\ref{conj:hyp2} is equivalent to its restriction to those log-concave measures
with zero mean and identity covariance (since $D(f)$ is an affine invariant).
Applying the identity \eqref{eq:chain} to such measures,
\ben
D(f)= \sum_{i=1}^n D(f_i) + I(f) ,
\een
since the standard normal is a product measure.
The lower bound of Proposition~\ref{prop:1d} asserts that
$h(X) \geq C + \log \sigma$ for one-dimensional log-concave distributions;
thus each $D(f_i)$ is bounded from above by some universal constant.
Thus $D(f)$ being uniformly $O(n)$ is equivalent to $I(f)$ being uniformly $O(n)$.
\end{proof}

Observe that mimicking Proposition~\ref{prop:1d}, Conjecture~\ref{conj:hyp2} may be written in the form: 
for a log-concave random vector $X$ taking values in $\RL^n$, 
\be\label{form1}
\frac{1}{n}\, h(X) \geq C + \log \sigma,
\ee
or
\be\label{form2}
\frac{1}{n}\, h(X) \geq \frac{1}{n}\, h(Z)- C' ,
\ee
where $C,C'$ are universal constants, and  $Z$ is the normal with the same covariance matrix as $X$.
Owing to \eqref{strengthen}, the form \eqref{form1} would strengthen
the naive lower bound of Proposition~\ref{prop:lc}. 
As for form \eqref{form2}, it looks like the lower bound of Theorem~\ref{thm:gaus-comp},
except that the way in which the matching Gaussian is chosen is to match
the covariance matrix rather than the maximum density. 

Existing partial results on the slicing problem already give insight
into the closeness of log-concave measures to Gaussian measures. 
For many years, the best known bound  in the slicing problem for general bounded
convex sets, due to Bourgain \cite{Bou91} in the centrally-symmetric case
and generalized by Paouris \cite{Pao00} to the non-symmetric case, 
was
\ben
L_K \leq  cn^{1/4}\log (n+1) .
\een
Recently Klartag \cite{Kla06} removed the $\log n$ factor and showed
that $L_K \leq cn^{1/4}$.  Using a transference result of Ball \cite{Bal88} from
convex bodies to log-concave functions, the same bound is seen to also apply 
to $L_f$, for a general log-concave density $f$.
Combining this with Corollary~\ref{cor:D-iso} leads immediately to the following
result.

\begin{prop}\label{prop:art}
There is a universal constant $c$ such that for  any log-concave density $f$ on $\RL^n$,
\ben
D(f)\leq \frac{1}{4} n\log n + cn .
\een
\end{prop}

Note that the property  \eqref{mot:1d} (quantified by Proposition~\ref{prop:1d}) 
for a one-dimensional log-concave density $f$ may be rewritten in the form
$0\leq D(f) \leq c$ for some constant $c$. 
Proposition~\ref{prop:art} is thus a multidimensional version of 
the statement \eqref{mot:1d}.


\section{Convexity of measures}
\label{sec:cvx}

Convexity properties of probability distributions may be expressed in terms 
of inequalities of the Brunn-Minkowski-type.
A probability measure $\mu$ on $\R^n$ is called $\k$-concave, where 
$-\infty \leq \k \leq +\infty$, if it satisfies
\be\label{defn-kconc}
\mu \big (tA + (1-t)B \big ) \geq
\big [\,t\mu(A)^\k + (1-t)\mu(B)^\k\big ]^{1/\k}
\ee
for all $t \in (0,1)$ and for all Borel measurable sets $A,B \subset \R^n$ 
with positive measure. Here $tA + (1-t)B = \{tx+(1-t)y: x \in A, y \in B\}$
stands for the Minkowski sum of the two sets.
When $\kappa = 0$, the inequality \eqref{defn-kconc} becomes
$$
\mu \big (tA + (1-t)B \big ) \geq \mu(A)^t \mu(B)^{1-t},
$$
and we arrive at the notion of a log-concave measure, introduced by 
Pr\'ekopa, cf. \cite{Pre71,Pre73,Lei72b}. In the absolutely continuous case,
the log-concavity of a measure is equivalent to the log-concavity of
its density, as in \eqref{defn:lc}. When $\kappa = -\infty$, the right-hand 
side is understood as $\min\{\mu(A), \mu(B)\}$. The inequality \eqref{defn-kconc} is 
getting stronger as the parameter $\k$ is increasing, so in the case 
$\k = -\infty$ we obtain the largest class, whose members are called convex 
or hyperbolic probability measures. For general $\k$'s, the family of 
$\k$-concave measures was introduced and studied by C. Borell \cite{Bor74, Bor75a}.

A remarkable feature of this family is that many important geometric 
properties of $\k$-concave measures, like the properties expressed in terms 
of Khinchin and dilation-type inequalities, may be controlled by the parameter 
$\k$, only, and in essence do not depend on the dimension $n$
(although the dimension may appear in the density
description of many $\k$-concave measures).

A full characterization of $\k$-concave measures was given by C. Borell in
\cite{Bor74, Bor75a}, cf. also \cite{BL76a}. Namely, any $\k$-concave probability measure is 
supported on some 
(relatively) open convex set $\Omega \subset \R^n$ and is absolutely 
continuous with respect to Lebesgue measure on $\Omega$. Necessarily,
$\k \leq 1/{\rm dim}(\Omega)$, and if $\Omega$ has dimension $n$, we have:

\begin{prop}\label{prop:kn-conc}
An absolutely continuous probability measure $\mu$
on $\R^n$ is $\k$-concave, where $-\infty \leq \k \leq 1/n$, if and only if
$\mu$ is supported on an open convex set $\Omega \subset \R^n$, where
it has a positive density $f$ such that, for all $t \in (0,1)$ and
$x,y \in \Omega$,
\be\label{kn-conc}
f(tx + (1-t)y) \geq
\big [\,t f(x)^{\k_n} + (1-t) f(y)^{\k_n} \big ]^{1/\k_n},
\ee
where $\k_n = \frac{\k}{1 - n\k}$.
\end{prop}

Following \cite{Avr72}, we call non-negative functions $f$, satisfying \eqref{kn-conc}, 
$\k_n$-concave. Thus, $\mu$ is $\k$-concave if and only if 
$f$ is $\k_n$-concave.

If $\k<0$, one may represent the density in the form
$f = \varphi^{-\beta}$ with $\beta \geq n$, $\k = - 1/(\beta - n)$,
where $\varphi$ is an arbitrary positive convex function on $\Omega$, satisfying
the normalization condition $\int_\Omega \varphi^{-\beta}\,dx = 1$.
Moreover, the condition $\beta \geq n+1$ like in Theorem~\ref{thm:k-conc-comp} corresponds to
the range $-1 \leq \k < 0$.

\begin{rmk}\label{rmk:pareto}
Note that a density of form $f = \varphi^{-\beta}$,
where $\varphi$ is an arbitrary positive convex function on $\Omega$
and $\beta < n$, need not be the density of a convex measure.
Indeed, it is not unless $\varphi$ itself can be written as a convex function 
raised to a large enough power. 
\end{rmk}

Proposition~\ref{prop:kn-conc} remains to hold without the normalization condition
$\mu(\R^n) = 1$. In particular, if $f$ is a positive $\k_n$-concave function 
on an open convex set $\Omega$ in $\R^n$, then the measure $d\mu(x) = f(x)\,dx$
is $\k$-concave, that is, it satisfies the Brunn-Minkowski-type inequality \eqref{defn-kconc}. 
For example, the Lebesgue measure on $\R^n$ is $\frac{1}{n}$-concave
(in which case $\k_n = +\infty$).

This sufficient condition will be used in dimension one as the following:

\begin{cor}\label{cor:1d-kconc}
Let $\alpha > 0$. If $u$ is a positive concave 
function on an interval $(a,b) \subset \R$, then the measure on $(a,b)$
with density $u^\alpha$ is $\frac{1}{\alpha + 1}$-concave.
\end{cor}


\section{Log-concavity of norms of convex functions}
\label{sec:norms}

In order to present the proof of our main result for $\kappa$-concave probability measures
(which we will do in Section~\ref{sec:proof2}),
we first need to develop some functional-analytic preliminaries.
 
Given a measurable function $f$ on a measurable set $\Omega \subset \R^n$
(of positive measure), we write
$$
\|f\|_p = \bigg(\int_\Omega |f|^p\,dx\bigg)^{\!1/p}, \quad -\infty<p<+\infty.
$$
For the value $p=0$, the above expression may be understood as the geometric mean
$\|f\|_0 = \exp \int \log |f|\,dx$.

It is easy to see that the function $p \rightarrow \|f\|_p^p$ is log-convex
(which is referred to as Lyapunov's inequality).
C. Borell complemented this general property with the 
following remarkable observation (\cite[Theorem 2]{Bor73a}). 

\begin{prop}\label{prop:borell-lc}
If $\Omega$ is a convex body, and if $f$ is positive and concave on $\Omega$, 
then the function
\be\label{borell}
p \longrightarrow C_{n+p}^{n}\, \|f\|_p^p\ = \frac{(p+1)\dots(p+n)}{n!}\
\int_\Omega f^p\,dx
\ee
is log-concave for $p>0$. 
\end{prop}

Here we use the standard binomial coefficients
\ben
C_{q}^{n} = \frac{q(q-1)\dots(q-n+1)}{n!} .
\een

Borell's theorem, Proposition~\ref{prop:borell-lc}, may formally be generalized to the class of
$\k$-concave functions $f$ with $\k>0$, since then $f = \varphi^{\k}$ 
with concave $\varphi$, and one may apply the log-concavity result \eqref{borell}, 
as well as the inequality \eqref{berwald} to $\varphi$.
However, for the purpose of proving Theorem~\ref{thm:k-conc-comp}, with the aim
of going beyond log-concave probability measures,
we are mostly interested in the case where $\k<0$, when the function 
$\varphi$ is convex. 

Thus what we require is a version of Proposition~\ref{prop:borell-lc} for convex functions $\phi$.
The following theorem, proved in \cite{BM10:norm}, supplies such a result.

\vskip5mm
\begin{thm}\label{thm:re-borell}
If $\varphi$ is a positive, convex function
on a open convex set $\Omega$ in $\R^n$, then the function
\be\label{re-borell}
p \longrightarrow C_{p-1}^{n}\, \|\varphi\|_{-p}^{-p}\ = 
\frac{(p-1)\dots(p-n)}{n!}\ \int_\Omega \varphi^{-p}\,dx
\ee
is log-concave on the half-axis $p>n+1$. 
\end{thm}

It is interesting to note that Borell \cite{Bor73a} obtained a different proof of Berwald's
inequality \cite{Ber47}, which is famous among functional analysts,
as a consequence of Proposition~\ref{prop:borell-lc}.

\begin{prop}\label{prop:berwald}
For $0<p<q$,
\be\label{berwald}
\left(C_{n+q}^{n}\,|\Omega|^{-1}\right)^{1/q}\ \|f\|_q \leq
\left(C_{n+p}^{n}\,|\Omega|^{-1}\right)^{1/p}\ \|f\|_p .
\ee
Equality is achieved when the normalized norms
are constant, which corresponds to the linear function 
$f(x) = x_1 + \dots + x_n$ on the convex body 
$$
\Omega = \{x \in \R^n: x_i > 0, \ x_1 + \dots + x_n < 1\}.
$$
\end{prop}

Berwald's inequality turns out to have interesting applications to
information theory as well as convex geometry (see \cite{BM10:norm, BM10:repi}).


\section{Entropy of $\kappa$-concave distributions}
\label{sec:proof2}

In this section, we explain how to use the remarkable property
of convex functions described by Theorem~\ref{thm:re-borell} in proving 
Theorem~\ref{thm:k-conc-comp}. 

\begin{proof}{\bf (of Theorem~\ref{thm:k-conc-comp}.)}  
Let $f = \varphi^{-\beta}$ be a probability density for a random vector
$X$ in $\R^n$ with $\beta \geq n+1$, where $\varphi$ is as in Theorem~\ref{thm:re-borell}.
Define $f$ to be zero outside $\Omega$.
As is shown in \cite{Bob07}, the density admits a bound
$$
f(x) \leq \frac{C}{(1 + |x|)^\beta}, \quad {\rm for \ all} \ \ x \in \Omega,
$$
with some constant $C$, so $\varphi(x) \geq c\,(1+|x|)$ with some $c>0$
(depending on $\varphi$). Hence, the function 
$$
V(p) = \log \int_\Omega \varphi^{-p}\,dx
$$
is finite and differentiable for $p>n$ with 
$V'(p) = - 
\int_\Omega \varphi^{-p}\,\log \varphi\ dx/\int_\Omega \varphi^{-p}\,dx$.
In particular,
\be\label{eq0}
h(X) = -\beta\, V'(\beta).
\ee

To proceed, assume $\|f\| = 1$, that is, $\inf_\Omega \varphi = 1$.
This assumption can be made since the quantity of interest in Theorem~\ref{thm:k-conc-comp},
$h(X)+\log \|f\|$ is an affine invariant; so one can scale $X$ to make $\|f\|=1$.
Like in the proof of Theorem~\ref{thm:lc}, for $t \in [0,1]$, write
$$
f(tx + (1-t)y) \geq
\left[\,t f(x)^{-1/\beta} + (1-t) f(y)^{-1/\beta}\right]^{-\beta},
$$
which is valid for any  $x,y \in \Omega$.
Integrating with respect to $x$ over the whole space, we get
$$
t^{-n} \geq \int 
\left[\,t f(x)^{-1/\beta} + (1-t) f(y)^{-1/\beta}\right]^{-\beta}\, dx
$$
with equality at $t=1$. Hence, we may compare derivatives of both sides
with respect to $t$ at this point,
which gives
$$
n\, \geq\, \beta \int
f(x)^{1 + 1/\beta} \left[f(x)^{-1/\beta} - f(y)^{-1/\beta}\right]\, dx.
$$
Optimizing over all $y$'s and using $\int f\,dx = 1$, we arrive 
at the bound
\be\label{eq1}
n \geq \beta \left(1 - \int_\Omega \varphi^{-\beta-1}\,dx\right).
\ee
Note that for the Pareto distribution there is equality at every
step made before (since in that case $\varphi$ is affine).

Now, it is time to apply Theorem~\ref{thm:re-borell}. Put 
$U(p) = \log \big[(p-1)\dots(p-n)\big]$, so that $R(p) = U(p) + V(p)$ 
is concave on the half-axis $p \geq n+1$. The concavity implies that
$$
R'(\beta) \geq R(\beta+1) - R(\beta).
$$
Equivalently, since $V(\beta) = 0$ and 
$U(\beta+1) - U(\beta) = \log \frac{\beta}{\beta - n}$, we have
\be\label{eq2}
V'(\beta) \geq V(\beta+1) + \log \frac{\beta}{\beta - n} - U'(\beta).
\ee
But \eqref{eq1} is telling us that $V(\beta+1) \geq \log(1 - \frac{n}{\beta})$,
so by \eqref{eq2},
$$
V'(\beta) \geq - U'(\beta) = -\sum_{i=1}^n \frac{1}{\beta - i}.
$$
With the representation \eqref{eq0} we arrive at the bound \eqref{kconc-ub},
$$
h(X) \leq \beta \sum_{i=1}^n \frac{1}{\beta - i}.
$$
From Lemma~\ref{lem:pareto2}, this is recognized as the entropy of the $n$-dimensional
Pareto density \eqref{gpd}, and hence Theorem~\ref{thm:k-conc-comp}
describes an extremal property of the Pareto distribution.
\end{proof}
\vspace{.1in}

In fact, as done for log-concave distributions in Section~\ref{sec:proof1},
it is possible obtain analogous bounds for the R\'enyi entropy
of any order.
We only state the result here; it is proved in \cite{BM10:norm}.

\begin{thm}\label{thm:renyi-cvx}
Fix $p\in (1,\infty)$. If a random vector $X$ in $\R^n$ has  density $f$ 
and a $\k$-concave distribution for $-1\leq\k<0$, then
\ben\begin{split}
0 \,&\leq\, \nth h_p(X) -  \log\, \|f\|^{-1/n}\\
 &\leq\, \frac{1}{p-1}\log \bigg[ \frac{(\beta p-1)\ldots(\beta p-n)}{(\beta-1)\ldots(\beta-n)} \bigg] ,
\end{split}\een
where $\beta=n+\frac{1}{(-\k)}$. 
\end{thm}

Consequently one has an extension of Corollary~\ref{cor:renyi-compare}
to the convex measure case:  for $X$ as in Theorem~\ref{thm:renyi-cvx}
and any $1<p< q \leq \infty$,
\ben\begin{split}
0 &\leq \frac{h_p(X)}{n}-\frac{h_q(X)}{n} \\
&\leq\, \frac{1}{p-1} \sum_{i=1}^n \log \bigg[1+ \bigg(\frac{\beta}{\beta-i}\bigg) (p-1) \bigg] .
\end{split}\een

To conclude this section, we show how Theorem~\ref{thm:k-conc-comp}
implies Corollary~\ref{cor:beta}.

\begin{proof}{\bf (of Corollary~\ref{cor:beta}.)}  
For Corollary~\ref{cor:beta}, observe that in the regime $\beta\geq \beta_0 n$,
\ben\begin{split}
\frac{\beta}{n} \sum_{i=1}^n \frac{1}{\beta - i}
&\leq \frac{1}{n} \sum_{i=1}^n \frac{1}{1- \frac{i}{\beta_0 n}} 
= \beta_0 \sum_{i=1}^n \frac{1}{\beta_0 n-i} \\
&\leq \beta_0 \frac{n}{\beta_0 n-n} 
= \frac{\beta_0}{\beta_0-1} .
\end{split}\een
On the other hand, in the regime $\beta\geq \beta_0 + n$,
\ben\begin{split}
\frac{\beta}{n} \sum_{i=1}^n \frac{1}{\beta - i}
&\leq \frac{\beta_0+n}{n} \sum_{i=1}^n \frac{1}{\beta_0 +n-i} \\
&\leq \frac{\beta_0+n}{n} \bigg[ \frac{1}{\beta_0 +n-1} + \log \bigg( \frac{\beta_0 +n-1}{\beta_0}\bigg) \bigg] \\
&\leq \frac{1}{n-1}+ \bigg(1+\frac{\beta_0}{n} \bigg)  \log \bigg( 1+\frac{n-1}{\beta_0}\bigg)  ,
\end{split}\een
as long as $\beta_0>0$. This gives an explicit bound for the $O(\log n)$
term in the second part of Corollary~\ref{cor:beta}.
\end{proof}


\section{Applications}
\label{sec:appln}

\subsection{Entropy rates}
\label{ss:rate}

Our first application is to the entropy rate of (strongly) stationary log-concave
random processes. We call a discrete-time stochastic process ${\bf X}=(X_i)$ 
{\it log-concave} if all its finite-dimensional marginals are log-concave distributions.
In particular, for the process ${\bf X}$ to be log-concave, it is necessary
and sufficient for the distribution of $X^n=(X_1,\ldots,X_n)$ to be log-concave for each $n$.
Note that an important special case of a log-concave process is a Gaussian process.

An important functional of a discrete-time stochastic process is its {\it entropy rate},
which is defined by
\ben
h({\bf X})=\lim_{n\ra\infty} \frac{h(X^n)}{n} ,
\een
when the limit exists. 

The only class of processes for which the computation of entropy rate is 
tractable is the class of stationary Gaussian processes.
Indeed, a stationary zero mean Gaussian random process is completely described 
by its mean correlation function $r_{k,j} = r_{k-j} = E[X_k X_j]$ 
or, equivalently, by its power spectral density function $G$, the Fourier transform of the covariance function:
\ben
G(\lam)=\sum_{n\in\IN} r_n e^{in\lam} .
\een
For a fixed positive integer $n$, the probability density function of $X^n$ is the normal density with 
$n\times n$ covariance matrix $R_n$, whose entries are $r_{k,j} = r_{k-j}$, and its entropy can be
explicitly written. 
This yields
\ben
h({\bf X})= 
\half \log(2\pi e) + \lim_{n\ra\infty} \nth \log \det (R_n) .
\een
Since $R_n$ is the Toeplitz matrix  generated by the power spectral density $G$ (or equivalently by the
coefficients $\{r_n\}$), 
one has from the theory of Toeplitz matrices (see, e.g., (1.11) in Gray \cite{Gra06})
that 
\ben
h({\bf X})=\half \log(2\pi e)+\frac{1}{2\pi} \int_0^{2\pi} \log G(\lam) d\lam .
\een
Below we point out that our inequalities give a way of obtaining some information
about the entropy rate of a stationary log-concave process.

\begin{cor}\label{cor:entrate}
For any stationary process ${\bf X}$ whose finite dimensional marginals
are absolutely continuous with respect to Lebesgue measure, let 
$f_n$ be the joint density of $X^n$. If 
\ben
f_{-} :=\liminf_{n\ra\infty} \nth \log\, \|f_n\|^{-1} >-\infty ,
\een
then the entropy rate $h({\bf X})$ exists and $h({\bf X})>-\infty$.
If, furthermore, ${\bf X}$ is a log-concave process and
\ben
f_{+} :=\limsup_{n\ra\infty} \nth \log\, \|f_n\|^{-1} <+\infty ,
\een
then
\ben
h({\bf X})  \leq\, f_{+} + 1.
\een
\end{cor}

\begin{proof}
Let $h(X|Y)$ denote conditional entropy.
As is well known (see, e.g., Cover and Thomas \cite{CT91:book}),
for any stationary process ${\bf X}$, the sequence
$a_n:= h(X_n|X^{n-1})$ is a non-increasing sequence, since
\ben\begin{split}
h(X_n|X^{n-1}) &\leq h(X_n|X_2, \ldots, X_{n-1}) \\
&=h(X_{n-1}|X^{n-2}) .
\end{split}\een
(Here one uses the fact that conditioning cannot increase entropy,
and the assumed stationarity.) Note also that
\ben
b_n:=\nth h(X^n) =\nth\sum_{i=1}^n a_i .
\een
Since 
\ben
\liminf_{n\ra\infty} b_n \geq \liminf_{n\ra\infty} \nth \log\, \|f_n\|^{-1}
=  f_{-} >-\infty ,
\een
we must have $\liminf_{n\ra\infty} a_n >-\infty$, which combined
with the monotonicity of $a_n$ implies that the limit exists and is 
equal to some $a>-\infty$.
Hence the limit of $b_n$, namely the entropy rate, also 
exists and is equal to $a$. The upper bound for the entropy rate
follows from Proposition~\ref{prop:lc}. 
\end{proof}

One interesting class of processes where this result may be of utility,
and where the study of entropy rate has attracted much recent interest,
is the class of hidden Markov processes.

Let us also note that reasoning similar to that in Corollary~\ref{cor:entrate} 
can be applied to bound the entropy rate of
continuous-time stationary log-concave processes as well (modulo some 
additional technicalities).

\subsection{The behavior of maximum density on convolution}
\label{ss:conv}

Our Proposition~\ref{prop:lc} can be used to significantly
generalize and improve an inequality of Junge \cite{Jun95} for the 
behavior of the maximum of a density on convolution.

\begin{cor}\label{cor:junge}
Let $f$ be the density of a $\kappa$-concave measure on $\RL^n$, where $\kappa\in[-1,0]$.
Then, for any $m\in\Nat$,
\ben
\|f^{*m}\| \leq \bigg(\frac{e^{1-\kappa n}}{\sqrt{m}}\bigg)^n \|f\| .
\een
\end{cor}
\begin{proof}
We wish to apply Corollary~\ref{cor:beta}, which requires
$\beta\geq \max\{\beta_0 n, n+1\}$ for some $\beta_0>1$.
Since $f$ is the density of a $\kappa$-concave measure,
it is of the form $\varphi^{-\beta}$ with $\varphi$ convex,
with $\kappa(\beta-n)=-1$ (see Section~\ref{sec:cvx}).
Thus the optimal (dimension-dependent!) $\beta_0$ that can be chosen in applying
Corollary~\ref{cor:beta} is given by
\ben
\kappa=\frac{-1}{\beta-n}
\geq \frac{-1}{\max\{1, n(\beta_0-1)\}}
= -\min\bigg\{1, \frac{1}{n(\beta_0-1)} \bigg\} ;
\een
in other words, one may take $\beta_0=1+(-\kappa n)^{-1}$,
for which one has
$C_{\beta_0}=\frac{\beta_0}{\beta_0-1}=1-\kappa n$.
Now if $X_i\sim f$ are i.i.d., and $S_m=\sum_{i=1}^m X_i$,
then 
\ben\begin{split}
nC_{\beta_0}+ \log \|f^{*m}\|^{-1} &\geq  h(S_m) \\
&\geq h(X_1) + \frac{n}{2}\log m \\
&\geq  \log \|f\|^{-1} + \frac{n}{2}\log m ,
\end{split}\een
by using Corollary~\ref{cor:beta},
the Shannon-Stam entropy power inequality \cite{Sta59},
and the first part of Proposition~\ref{prop:lc}. Exponentiating yields
the desired result.
\end{proof}

In particular, for a log-concave density $f$ on $\RL^n$,
\ben
\|f^{*m}\| \leq \bigg(\frac{e}{\sqrt{m}}\bigg)^n \|f\| .
\een
While Junge \cite{Jun95} proved that for a symmetric, log-concave density $f$,
\be\label{junge}
\|f^{*m}\| \leq \bigg(\frac{c}{\sqrt{m}}\bigg)^n \|f\| 
\ee
for some universal constant $c$,
Corollary~\ref{cor:junge} above generalizes this by removing the symmetry
assumption, making the universal constant explicit, and slightly broadening
the class of densities allowed; also the proof is far more elementary.

Let us observe that in the three inequalities in the proof of Corollary~\ref{cor:junge},
one is tight only for uniforms on convex sets, another only for Gaussians, and the 
third only for Pareto-type distributions; so Corollary~\ref{cor:junge} is always loose,
although it is possible that $c=e$ could be the optimal dimension-free constant in \eqref{junge}.

\subsection{Infinitely divisible distributions}
\label{ss:ID}

Our third application is to estimating the entropy of certain infinitely divisible distributions.
Let $X$ be a random vector in $\R^n$ with density $f(x)$ and
characteristic function
$$
\varphi(t) = \E\, e^{i \left<X,t\right>} = 
\int e^{i \left<x,t\right>}\, f(x)\, dx.
$$
Recall that the distribution of $X$ is {\it infinitely divisible} if $X$ can be realized as
the sum of $M$ independent random vectors, for any natural number $M$.
The L\'{e}vy-Khintchine representation theorem \cite{Sat99:book,App04:book} asserts that
the distribution of $X$ is infinitely divisible if and only if
there exist a symmetric nonnegative-definite $n\times n$ matrix $\Sigma$,
$\gamma\in\R^{n}$ and a L\'evy measure $\nu$
such that the characteristic function $\varphi(t)$ of $X$ is given by
\ben\label{LevyKhintRep}\begin{split}
\varphi(t)=\exp\bigg\{-&\frac{1}{2} \langle\Sigma t, t\rangle
\!+\mathrm{i} \langle\gamma,t\rangle\\
&  +\int_{\R^{n}}\left(  \mathrm{e}^{\mathrm{i}\langle x,t\rangle
}\!-1- \frac{\mathrm{i}\langle x.t\rangle}{1+\left\vert x\right\vert ^{2}%
}\right)  \nu(\mathrm{d}x)\bigg\},
\end{split}\een
for each $t\in\R^{n}$.
Here,  a measure $\nu$ on $\R^{n}$ is called a {\it L\'evy measure} 
if it satisfies $\nu(\{0\})=0$ and $\int_{\R^n}(1\wedge\left\vert x\right\vert ^{2})\nu(dx)<\infty. $ 
The triplet $(\Sigma,\nu,\gamma)$ is called the L\'{e}vy-Khintchine triplet of 
$X$. We write $ID(\Sigma,\nu,\gamma)$ for the distribution of $X$, and use the
abbreviation $ID(\nu):=ID(0,\nu,0)$. Fixing $\gamma=0$ is just fixing a location
parameter and does not matter for the entropy, whereas fixing $\Sigma=0$
means that the infinitely divisible measure has no Gaussian part.

We start with a one-dimensional result.

\begin{cor}
Let $\nu$ be any log-concave \levy measure supported on $(0,\infty)$. Assume that the density 
$m(x)$ of $\nu$  satisfies $m(0+)\geq 1$.
Then if $f$ is the density of $ID(\nu)$, 
\ben
h(ID(\mu)) \leq 1-\log \|f\| .
\een
\end{cor}

\begin{proof}
Yamazato \cite{Yam82} (see also Hansen \cite{Han88} for an alternative proof) showed that
for infinitely divisible measures supported on the positive real line,
if the \levy measure has a log-concave density $m$, then the density of 
the ID measure is log-concave if and only if $m(0+)\geq 1$.
\end{proof}

\medskip
It is natural to ask how to bound the entropy $h(X)$ in terms of
$\varphi$, especially when $f$ is not given explicitly but 
$\varphi$ is (which is typical in the case of infinitely divisible distributions). 
We show below that some explicit bounds may be given when 
we know something about convexity properties of the density.
The idea is to utilize the R\'enyi entropy of order 2, 
since it is directly connected to the characteristic function by Plancherel's
identity.

To start with, assume $f$ is log-concave. Then by applying
Corollary~\ref{cor:renyi-compare} with $p=1$ and $q=2$, one obtains
\ben
-\log\, \|f\|_2^2 \leq\, h(X)  \leq\, n - \log\, \|f\|_2^2 ,
\een
since by definition, $h_2(X)=- \log\, \|f\|_2^2$.
Note that the lower bound here is 
universally true (for all random vectors $X$) as a consequence of Jensen's
inequality; indeed, if $X$ has density $f$,
\ben
h(X)=E[-\log f(X)] \geq -\log Ef(X) = h_2(X).
\een

But Plancherel's formula asserts that $\|f\|_2^2 = (2\pi)^{-n}\, \|\varphi\|_2^2$.
Hence:

\begin{prop}\label{prop:chfn}
Let $X$ be a log-concave random vector in $\R^n$ 
with characteristic function $\varphi(t)$. Then
\ben
n \log(2\pi) -\log\, \|\varphi\|_2^2 \leq\, h(X)  \leq\, 
n \log(2\pi e) - \log\, \|\varphi\|_2^2,
\een
where
$$
\|\varphi\|_2^2 = \int |\varphi(t)|^2\,dt.
$$
\end{prop}

Equivalently,
$$
\log(2\pi) \, \leq\, \frac{1}{n}\,h(X) + \|\varphi\|_2^{2/n}  \leq\, \log(2\pi e).
$$
This gives a reasonably strong approximation for the
entropy of a log-concave distribution that is only known through its
characteristic function: the gap between the upper and lower bounds is just 1.


One would also hope to be able to bound the entropies of the non-normal stable laws 
(which are not log-concave). As a step in this direction, 
we have a generalization of Proposition~\ref{prop:chfn} to the $\k$-concave case.

\begin{thm}\label{thm:chfn-k}
If $X$ has a $\k$-concave distribution, 
\be\label{ID-want}\begin{split}
h_2(X)
\leq h(X) 
\leq  h_2(X)+ \beta \sum_{i=1}^n \frac{1}{\beta - i} ,
\end{split}\ee
provided $\beta=n-\frac{1}{\k} \geq n+1$.
\end{thm}

The upper bound is easy to see using Theorem~\ref{thm:k-conc-comp} 
(more precisely, inequality \eqref{kconc-ub})
and the first part of Theorem~\ref{thm:lc};
the lower bound is as for Proposition~\ref{prop:chfn}.

Observe that $h_2(X)$ can be explicitly computed in many interesting cases via Plancherel's formula.
For instance, for {\it one-dimensional} symmetric $\alpha$-stable measures with characteristic function
\ben
\varphi(t)=\exp(-|t|^\alpha) ,
\een
one obtains
\ben\begin{split}
\|\varphi\|_2^2 &= 2 \int_0^\infty  \exp(-2t^\alpha) dt \\
&= \frac{1}{\alpha} \int_0^\infty \bigg(\frac{z}{2}\bigg)^{\frac{1-\alpha}{\alpha}} e^{-z} dz \\
&= \frac{2^{1-\frac{1}{\alpha}}}{\alpha} \Gamma\bigg(\frac{1}{\alpha}\bigg) .
\end{split}\een

For the sake of illustration, let us apply our inequalities to approximating 
the entropy of the Cauchy  distribution, which is also explicitly computable.
Recall that the standard Cauchy distribution (stable index $\alpha=1$, skewness parameter $\beta=0$) has density 
\ben
f(x)=\frac{1}{\pi(1+x^2)} ,   x\in \RL ,
\een
and entropy $\log (4\pi)\approx 1.386 +\log\pi$.
It is easy to check that the Cauchy distribution is $-1$-concave (i.e., one can choose
$\beta=2$), so applying Theorem~\ref{thm:k-conc-comp} gives
\ben
h(X)\leq \log \pi + 2
\een
since $\|f\|=1/\pi$. On the other hand,
$h(X)\geq h_2(X)= -\log [(2\pi)^{-1}\Gamma(1)]= \log (2\pi)$,
so that $h(X)$ is trapped in a range of width $2-\log 2\approx 1.307$
centered at approximately $1.347 +\log\pi$,
which seems fairly good.

For {\it multivariate} symmetric $\alpha$-stable probability measures,
a representation of the R\'enyi entropy of order 2 is obtained by Molchanov \cite{Mol09},
in terms of the volume of a star body associated with the measure. In particular, 
\cite{Mol09} uses the identity 
\ben
\int_{\RL^n} f(x)^2 dx =2^{-n/\alpha} f(0)
\een
for any symmetric $\alpha$-stable density $f$; as pointed out by a reviewer,
the left side here is just the value of the self-convolution of $f$ at 0,
and for $X, X'$ drawn independently from $f$, stability implies that
$X+X'$ has the same distribution as  $(1^\alpha+1^\alpha)^{n/\alpha} X$.
It is well known that symmetric stable random vectors
are unimodal with mode at 0 (see, e.g., Kanter \cite{Kan77}); hence one can rewrite this as
\ben
\nth h_2(X)=\log \|f\|^{-1/n} +\frac{1}{\alpha} \log 2.
\een
However, this still does not seem as useful as using Plancherel's formula
to connect with the characteristic function.

It seems plausible that large classes of infinitely divisible distributions
are $\k$-concave, although we do not know any existing general results in this direction (other than
the log-concavity results mentioned earlier). If one were able to get estimates on $\k$
for an infinitely divisible distribution that is specified through its characteristic function,
\eqref{ID-want} would immediately yield an upper bound for entropy in terms of  
$\|\varphi\|_2^2$.

Some negative results on $\k$-concavity of stable laws actually follow from the
preceding discussion. Indeed, if $X$ is symmetric $\alpha$-stable and $\k$-concave,
one has
\ben\begin{split}
\log \|f\|^{-1/n} +\frac{1}{\alpha} \log 2
&=\nth h_2(X)
\,\leq\nth h(X) \\
&\leq \log \|f\|^{-1/n} + \frac{\beta}{n} \sum_{i=1}^n \frac{1}{\beta - i} \\
&\leq \log \|f\|^{-1/n} + (1-\k n) ,
\end{split}\een
where the last inequality follows from a similar calculation
as in the proof of Corollary~\ref{cor:junge}.
Thus one obtains
\ben
\k \leq \bigg(1-\frac{\log 2}{\alpha}\bigg) \nth .
\een
This is rather loose, since we already know that for $\alpha< 2$, no symmetric $\alpha$-stable distribution
can be log-concave (as it would otherwise have finite moments). However, it does give some 
negative information for $\alpha<\log 2$. For instance, it shows that for fixed dimension $n$,
symmetric $\alpha$-stable distributions cease to be $\k$-concave for any fixed $\k\in (-\infty,0)$
as $\alpha\ra 0$. This leads us to the following conjecture. 

\begin{conj}\label{conj:stable}
Any strictly stable probability measure on an infinite-dimensional separable
Hilbert space is convex. In the finite-dimensional case, one has a threshold
phenomenon: For fixed $\k\leq 0$, a spherically symmetric stable 
distribution of index $\alpha$ on $\RL^n$
is $\k$-concave if and only if $\alpha\geq\alpha^*(\k,n)$, where $\alpha^*(\k,n)\in (0,2]$
is a constant depending only on $\k$ and $n$.
\end{conj}

Recall that a random element
$X$ in a separable Hilbert space is said to have a strictly stable distribution with index $\alpha\in (0,2]$
if $X_1+\ldots+X_m \eqD m^{1/\alpha} X$, where the $X_i$ are independent random elements
with the same distribution as $X$.

\subsection{Entropy of Mixtures}
\label{ss:mix}

Our fourth application is focused on estimating the entropy of scale mixtures of Gaussians
(or more generally log-concave distributions). Such distributions are of great interest
in Bayesian statistics. 

Suppose one starts with a log-concave density $f=e^{-\phi}$, where
$\phi$ is convex. A scale mixture using a mixing distribution with density $m$ on the positive
real line would have the density
\ben
\int_0^{\infty} m(s) \frac{1}{s} \exp\bigg\{-\phi\bigg(\frac{x}{s}\bigg)\bigg\} ds .
\een
More generally, one can consider ``multivariate scale mixtures'' of form
\ben
f_{\text{mix}}(x)=\int_{P(n)} m(A) f_A(x) \eta(dA) ,
\een
where 
\ben
f_A(x)= \frac{f(A^{-1}x)}{\det(A)} 
\een
is the density of $AX$ when $X$ is distributed according to $f$,
and $\eta$ represents the restriction of the 
Haar measure on the general linear group $GL(n)$ equipped with the
multiplicative operation to the  
subset $P(n)$ (which is both a semigroup with respect to matrix multiplication
and a cone, but not a group) of positive-definite matrices.

Note that lower bounds on entropy of mixtures are easy to obtain
by using concavity of entropy, 
but upper bounds are in general difficult.
Indeed,
\ben\begin{split}
h(f_{\text{mix}}) & \geq \int_0^{\infty} m(A)h(f_A) \eta(dA) \\
&= \int_0^{\infty} m(A) [h(f) +\log\det(A) ] \eta(dA) \\
&= h(f) + \int_0^{\infty} m(A) \log\det(A) \eta(dA) .
\end{split}\een
On the other hand, one has an upper bound under a log-concavity assumption.

\begin{thm}\label{thm:mix}
Suppose $f_{\text{mix}}$ is a scale mixture of the log-concave density $f=e^{-\phi}$, using
a mixing distribution with density $m$ on the positive-definite cone $P(n)$. Assume
$f$ has a mode at 0 (which for instance is the case when it is 
symmetric), and that $f_{\text{mix}}$ is log-concave. Then 
\ben
h(f_{\text{mix}}) \leq n+\phi(0)-\log \int_{P(n)} \frac{m(A)}{\det(A)} \eta(dA) .
\een
\end{thm}

\begin{proof}
The proof is obvious from Proposition~\ref{prop:lc} and 
the fact that the mixture density must also have its
mode at 0.
\end{proof}
\vspace{.1in}

The condition on Theorem~\ref{thm:mix} that the mixture be log-concave
may not be too onerous to check, at least in the case of mixtures involving
a one-dimensional scale, i.e., $A=sI$, with $m$ now a prior on $\RL_+$.
A sufficient condition for $f_{\text{mix}}$ to be log-concave is obtained by requiring
that the integrand above is log-concave (thanks to the Pr\'ekopa-Leindler inequality),
which means 
\ben
\bar{\phi}(x,s):=\phi\bigg(\frac{x}{s}\bigg) - \log\frac{m(s)}{s}
\een
is convex. For given $m$, this may be checked by 
verifying positive-definiteness of the $(n+1)\times (n+1)$ Hessian 
matrix of $\bar{\phi}$ for $x\in\RL^n, s\in(0,\infty)$.

One has an even simpler statement for Gaussian mixtures, which already appears
to be new.

\begin{cor}\label{cor:mix}
Suppose the mixing distribution with density $m$ on the positive real line
is slightly stronger than log-concave,
in the sense that 
$\frac{n}{2}\log v-\log m(v)$ is convex.
Writing $Z$ for a standard Gaussian on $\RL^n$ with density $g$, let the random vector 
$Y=\sqrt{V}Z$, where $V$ is a scalar distributed according to $m$, 
have the density $g_{\text{mix}}$.
Then
\ben\begin{split}
\frac{n}{2} \int_{0}^{\infty} m(v)\log v \, dv
&\leq  h(g_{\text{mix}})-h(g) \\
&\leq \frac{n}{2}-\log\int_{0}^{\infty} \frac{m(v)}{v} dv .
\end{split}\een
\end{cor}

\begin{proof}
One can write
\ben
g_{\text{mix}}(x)=\int_0^{\infty} \frac{m(v)}{(2\pi v)^{n/2}} \exp\bigg\{-\frac{\|x\|^2}{2v}\bigg\} dv ,
\een
since we are parametrizing using variance (rather than standard deviation).
Also, $\|x\|^2/v$ is convex as a function of $(x,v)\in \RL^n\times (0,\infty)$;
indeed, the quadratic form induced by its Hessian matrix when evaluated at
$(a,b)\in \RL^n\times (0,\infty)$ is $\|ya-bx\|^2\geq 0$. 
Combining this with the assumed log-concavity of  $\frac{m(v)}{v^{n/2}}$,
one finds that the integrand above is log-concave,
and hence so is $g_{\text{mix}}$ (by the Pr\'ekopa-Leindler inequality).
Then the first inequality of Corollary~\ref{cor:mix} follows from concavity
of entropy, while the second follows from  Proposition~\ref{prop:lc}.
\end{proof}

A limitation (perhaps unavoidable) of this result is that as dimension increases, the shape
requirement on the prior $m$ becomes increasingly stringent.


\section{Discussion}
\label{sec:disc}

A central result in our development was the identification of the maximizer
of R\'enyi entropy under log-concavity and supremum norm constraints.  
We gave a number of probabilistic, information theoretic and convex geometric 
motivations for considering this entropy 
maximization problem.

There are some other works in which both log-concavity and entropy appear, 
although they are only tangentially related to the substance of this paper.
Log-concavity plays a role in a few other entropy bounding problems-- see, for instance, 
Cover and Zhang \cite{CZ94} and Yu \cite{Yu08:kr}. 
Log-concavity (in the discrete sense) also turns out to be relevant to the behavior of discrete entropy; 
see Johnson \cite{Joh07} and 
\cite{JKM09:maxent} 
for examples. For instance, Johnson \cite{Joh07} showed that the Poisson is maximum entropy among all
ultra-log-concave distributions on the non-negative integers with fixed mean
(ultra-log-concavity is a strengthening of discrete log-concavity).

For completeness, let us also mention that a different maximum entropy characterization 
of one-dimensional generalized Pareto distributions
was given by Bercher and Vignat \cite{BV08:isit}. 
However, their characterization is rather different: 
in particular, they use R\'enyi and Tsallis entropies rather than Shannon entropy,
and also it is not clear what the motivation is for the somewhat artificial moment and 
normalization constraints they impose.
While \cite{BV08:isit} claims a connection to the Balkema-de Haans-Pickands theorem
for limiting distribution of excesses over a threshold, log-concavity does not play
a role in their development.


Our main goal in this paper was to better understand the behavior of entropy
for log-concave (and more generally, hyperbolic) probability measures,
particularly as regards phenomena that do not degrade in high dimensions. 
The information-theoretic perspective on convex geometry suggested in this paper 
appears to be bearing fruit; for instance, in \cite{BM10:repi}, we use some
of the results in this paper as one ingredient (among several) to prove a
``reverse entropy power inequality for convex measures'' analogous to Milman's
reverse Brunn-Minkowski inequality \cite{Mil86,Mil88:1, Mil88:2, Pis89:book} for convex bodies.

We conclude with some open questions. First, the question
of characterizing the $\kappa$-concavity properties of infinitely
divisible laws using only knowledge of the characteristic function
is an interesting one, as discussed in Section~\ref{ss:ID}. 
One also hopes that Theorem~\ref{thm:k-conc-comp}
can be improved to only require $\beta>n$; this would
immediately imply that many of the results in this paper stated
for $\kappa$-concave measures with $\kappa\geq -1$ (or their densities)
would have extended validity to general convex measures. 
And finally, it would be nice to use the entropic formulation
of the hyperplane conjecture to improve the state-of-the-art
partial results that exist.

\appendices
\section{The multivariate Pareto distribution}

There does not seem to be a canonical definition of a multivariate version for the 
Pareto distribution, although various versions appear to have been examined
in the actuarial literature (see, e.g., \cite{Mar62, Yeh04, AFV10}). For our purposes,
the distribution with density $f_{\beta,a}$ defined in \eqref{gpd} is the relevant
generalization. In this Appendix, we collect some simple observations about
this multivariate Pareto family. Recall that
\ben
f_{\beta,a}(x) = \frac{1}{Z_n(\beta,a)}\, (a + x_1 + \dots + x_n)^{-\beta}, \quad x_i>0 .
\een

\begin{lem}\label{lem:pareto1}
For any $a>0$, the normalizing factor 
$$
Z_n(\beta,a)= \int_{\RL_{+}^n} \,\, (a + x_1 + \dots + x_n)^{-\beta} \, dx
$$ 
is finite if and only if $\beta>n$. Moreover, for $\beta>n$,
$$
Z_n(\beta,a)= \frac{1}{(\beta-1)\ldots(\beta-n)} \cdot \frac{1}{a^{\beta-n}}.
$$ 
\end{lem}
\begin{proof}
We prove the desired statement by induction. First,
\ben\begin{split}
Z_1(\beta,a)&= \int_{0}^{\infty} \,\, (a + x)^{-\beta} \, dx 
= \int_{a}^{\infty} \,\, y^{-\beta} \, dy
= \frac{y^{1-\beta}}{1-\beta}\bigg|_{a}^{\infty} \\
&= 
\left\{ \begin{array}{lll} \infty & \textrm{if $\beta\leq 1$}\\ 
\frac{1}{\beta-1}\cdot\frac{1}{a^{\beta-1}} & \textrm{if $\beta>1$} 
\end{array} \right. 
\end{split}\een
Now assume that the statement is true for $Z_{n-1}$, and
observe that
\ben\begin{split}
Z_n(\beta,a)&= \int_{0}^{\infty} dx_n \int_{\RL_{+}^{n-1}} (a + x_1 + \dots + x_n)^{-\beta} \, dx_1\ldots dx_{n-1} \\
&= \int_{0}^{\infty}  \frac{1}{(\beta-1)\ldots(\beta-n+1)} \cdot \frac{dx_n}{(a+x_n)^{\beta-n+1}}   \\
&= \frac{1}{(\beta-1)\ldots(\beta-n+1)} Z_{1}(\beta-n+1,a) \\
&=  \frac{1}{(\beta-1)\ldots(\beta-n)} \cdot \frac{1}{a^{\beta-n}} ,
\end{split}\een
which is the required conclusion for $Z_n$.
\end{proof}
\vspace{.1in}

In particular, $f_{\beta,a}$ is not a well defined density for $\beta\leq n$,
and there is no Pareto distribution with such parameters.

\begin{lem}\label{lem:pareto2}
For any $a>0$ and $\beta>n$, the entropy of the 
multivariate Pareto distribution  $f_{\beta,a}$ is given by
\ben
h(f_{\beta,a}) + \log\, \|f_{\beta,a}\|^{-1}\, =\ \beta \sum_{i=1}^n \frac{1}{\beta - i}.
\een
\end{lem}
\begin{proof}
If $Y\sim f_{\beta,a}$, then
\ben
h(Y)= \log Z_n(\beta,a) + \frac{\beta}{Z_n(\beta,a)} L_n(\beta,a) ,
\een
where
\ben
L_n(\beta,a) = \int_{\RL_{+}^n} \frac{\log(a + x_1 + \dots + x_n)}{(a + x_1 + \dots + x_n)^{\beta}} dx .
\een
With this notation, what we wish to prove is that
\be\label{L-rec}
\frac{L_n(\beta,a)}{Z_n(\beta,a)} = \log a + \sum_{i=1}^n \frac{1}{\beta - i}.
\ee

As in the proof of Lemma~\ref{lem:pareto1}, one can write the recursion
\ben
L_n(\beta,a)= \int_{0}^{\infty} \,  L_{n-1}(\beta, a+x_n) \, dx_n ,
\een
and it is a simple exercise using integration by parts to see that
\be\label{eq:l1}
L_1(\beta,a)=Z_1(\beta,a) \bigg[\frac{1}{\beta-1}+\log a\bigg] .
\ee
Our goal is to prove the identity \eqref{L-rec} by induction.
To this end, we compute using the
induction hypothesis for $n-1$:
\ben\begin{split}
L_n(\beta,a)&= \int_{0}^{\infty} Z_{n-1}(\beta,a+y)\bigg[\log(a+y)+ \sum_{i=1}^{n-1} \frac{1}{\beta - i}\bigg] \, dy \\
&= \frac{1}{(\beta-1)\ldots(\beta-n+1)}  \int_{0}^{\infty} \frac{\log(a+y)}{(a+y)^{\beta-n+1}} dy \\
&\quad+ Z_{n-1}(\beta,a)\sum_{i=1}^{n-1} \frac{1}{\beta - i} .
\end{split}\een
Recognizing the integral in the last expression as $L_1(\beta-n+1,a)$
and plugging in the evaluation \eqref{eq:l1},
simple manipulations give us \eqref{L-rec}.
Observing that $\|f_{\beta,a}\|^{-1}= Z_n(\beta,a) a^{\beta}$, the proof
of Lemma~\ref{lem:pareto2} is complete.
\end{proof}

\section*{Acknowledgment}
We thank each of the 3 referees for a careful reading and many useful
comments.



\bibliographystyle{IEEEtranS}



\end{document}